\documentclass[11pt,twoside]{article}

\usepackage{asp2006}
\usepackage{epsf}
\usepackage{lscape}

\begin{document}
\title{Hot DQ White Dwarf Stars: A New Challenge to Stellar Evolution} 
\author{P. Dufour$^1$, J. Liebert$^1$, G. Fontaine$^2$, N. Behara$^3$} 
\affil{$^1$Department of Astronomy and Steward Observatory, University of Arizona, 933 North Cherry Avenue, Tucson, AZ 85721\\
$^2$D\'epartement de Physique, Universit\'e de Montr\'eal, CP6128, Succ. Centre-Ville, Montr\'eal, QC H3C 3J7, Canada\\
$^3$CIFIST, GEPI, Observatoire de Paris-Meudon, 92195, France}  

\begin{abstract} 

We report the discovery of a new class of hydrogen-deficient
stars: white dwarfs with an atmosphere primarily composed of
carbon, with little or no trace of hydrogen or helium. Our analysis
shows that the atmospheric parameters found for these stars do not fit
satisfactorily in any of the currently known theories of
post-asymptotic giant branch (AGB) evolution, although these objects
might be the cooler counter-part of the unique and extensively studied
PG 1159 star H1504+65. These stars, together with H1504+65, might thus
form a new evolutionary post-AGB sequence.
\end{abstract}


\section{Introduction} 

Canonical stellar evolution predicts that the majority of white dwarfs
have a core made of carbon and oxygen which itself is surrounded by a
helium layer and also, for $\sim$ 80 $\%$ of known white dwarfs, by an
additional hydrogen layer. Thus, all white dwarfs have been
traditionally found to belong to one of these two categories: those
with a hydrogen-rich atmosphere (the DAs) and those with a helium-rich
atmosphere (the non-DAs). The latter are believed to be the result of a
late He-shell flash \citep{herwig99} at the end of the post-AGB phase
(the so-called born-again scenario) which almost completely removes
hydrogen and mixes interior material (mostly carbon and oxygen) with
the helium envelope. With cooling, because of gravitational settling,
helium gradually diffuses upward and the stars then appear as
helium-rich white dwarfs (spectral type DO for $T_{\rm eff} >  \sim
40,000$ K and DB below that). When these white dwarfs cool to
approximately 13,000 K, carbon makes one final appearance as it is
convectively dredged-up from the diffusion tale by the deep superficial
helium convection zone \citep[DQ spectral
type,][]{pelletier86}. Evolutionary models \citep{fontaine05} predict
that the carbon abundance reaches a maximum at $\sim 10,000 $ K and
decreases monotonically with decreasing effective temperature (note
that even at maximum contamination, helium always remains the dominant
constituent of the atmosphere with log C/He $\sim -3$).

Detailed model atmosphere analysis of DQ white dwarfs \citep{dufour05}
have confirmed the theory on the cool side of the maximum contamination
but very little has been done to test the theory on the hot ascending
side of the curve. The reason for this is simple: only two objects
\citep[G227-5 and G35-26,][]{wegner85,thejll90} were known to exhibit
traces of carbon in the optical in that temperature
regime. Fortunately, the situation has recently changed drastically
thanks to the discovery of several new hot white dwarfs with carbon
lines in the Sloan Digital Sky Survey \citep{liebert03}. However,
appropriate atmospheric models were not available at the time, so
a detailed analysis of these stars could not be undertaken. We have thus
recently developed the appropriate atmospheric models and performed
the analysis of these objects.

\section{Spectroscopic and photometric analysis of hot DQ stars}

Most stars in our sample are very similar to the previously known DQ
stars G227-5 and G35-26 and exhibited mostly C\textsc{i} lines. However,
a few showed a more complex and rich spectrum that is attributed to
C\textsc{ii} absorption lines. While the former were relatively easy
to model (results will be presented elsewhere), it became rapidly
obvious that the latter could not be modeled by simply increasing the
effective temperature. Indeed, in such cases, helium is no longer
spectroscopically invisible and strong HeI lines ($\lambda$4471 in
particular) are expected. We soon came to the stunning conclusion that
no combination of carbon and helium could successfully reproduce the
observed features by assuming a helium-dominated atmophere. We instead
find that a good fit to both the spectra and energy distribution is
possible only by considering an atmophere made primarily of carbon,
with little or no trace of hydrogen and helium !! We found 8 such objects
in the SDSS DR4 archive, 6 of which are presented in figure 1. Since
the signal-to-noise ratio of the SDSS spectra is rather poor, only crude
upper limits (except for two that show explicitly H$\alpha$ and
H$\beta$) can be obtained for the hydrogen and helium abundances
(higher quality data should be secured shortly).

\section{H1504+65 and evolutionay scenarios}

How are these stars formed ? The most attractive scenario is that
these are the progenies of objects such as the unique hot PG1159 star
H1504+65 \citep{werner91,werner99,werner04}. The latter is
the only other known star not showing any trace of hydrogen and helium
in its spectra and has a very unusual composition with $\sim 50\%$ C
and $\sim 50\%$ O. Although the exact reason for the He-deficiency is
still not fully understood to this day, it is believed that H1504+65 is
essentially a bare stellar core that might be the result of a
particularly violent post-AGB very late thermal pulse which has
destroyed the remaining stellar envelope containing helium and
hydrogen. It could also be the result of a heavy-weight
intermediate-mass star (8 $M_{\odot} \leq$ M $\leq 10 M_{\odot}$) that
has gone through carbon burning and has a O-Ne-Mg core
\citep{werner91, werner99,ritossa99}. A critical test of this idea is
to measure the mass of the remnants since the offspring of such
massive stars are expected to be massives as well ($\geq$ 1
$M_{\odot}$). Unfortunately, the quality of the data at hand
does not allow a precise surface gravity measurement although it seems,
at least for some stars, that fits with models with high surface gravity have
lines that are much too broad. However, given the low signal-to-noise ratio
of the observations and the uncertainties in the broadening theory for
the carbon lines, we remain cautious and will refrain from drawing any
definitive conclusions at this point.

We will conclude with some thoughts on the spectral evolution of these
stars. All carbon-rich white dwarfs we have found so far have
effective temperatures between $\sim$ 18,000 and 24,000 K (depending
on our choice of surface gravity, abundances of trace elements such as
H, He and O and the amount of dereddening applied). Why do we find
carbon-rich objects only in that narrow range of effective temperature
? If they are descendants of a star like H1504+65 ($T_{\rm eff} \sim
200,000$ K), why don't we see carbon/oxygen-rich white dwarfs at
intermediate temperature? We believe that the simplest way to explain
this is that a star like H1504+65, however it was formed, most
probably still contains a tiny amount of helium which will eventually
diffuse upward to form a thin layer ($\sim 10^{-15} M_*$ is enough to
form a full atmosphere!)  above the C-enriched and O-depleted
mantle. Such stars would thus cool as normal He-rich white dwarfs
until the underlying carbon convection zone develops (due to its
recombination) and completely dilutes from below the tiny overlaying
He layer.  Hence, a star like H1504+65, showing initially a mixed C
and O atmosphere, would then temporarily disguise itself as a
helium-rich star before transforming itself into a carbon dominated
atmosphere star at an effective temperature whose exact value is not
yet known due to a current lack of proper models (hopefully around
$\sim$ 24,000 K !). Furthermore, the absence of carbon-rich white
dwarfs at low temperatures imply that these stars must again go
through a drastic spectral change as they cool. We speculate that
helium ultimately re-emerge at the surface, perhaps turning the
carbon-rich white dwarfs into DQ white dwarfs belonging to the second
sequence with higher carbon abundance \citep{dufour05,koester06}. In
conclusion, the surprising discovery of several white dwarfs stars
with a carbon-rich atmosphere, and the explanation of their origin and
evolution, represent new challenges that should ultimately deepen our
understanding of stellar evolution.


\acknowledgements 
This work has been partially supported by the NSF grant AST 03-07321 for 
work on the SDSS white dwarfs. This work was also supported in part by
the NSERC (Canada).


\end{document}